\documentstyle[12pt]{article} \long\def\del#1\enddel{ }
\let\3=\ss \catcode`\"=\active \let"=\"
  \oddsidemargin=\evensidemargin
\let\ssk=\smallskip  
  
\let\qd=\quad \let\qqd=\qquad  \def\ve{\vfil\eject}

\let\a=\alpha  \let\g=\gamma  \let\e=\varepsilon
\let\z=\zeta    
    \let\p=\pi \let\r=\rho
\let\s=\sigma \let\t=\tau  
 \let\ph=\varphi
\let\Ph=\Phi  \let\Ps=\Psi   
   
\def\0{\over }    \def\1{\vec }   \def\2{{1\over2}} \def\3{{\ss}}
\def\4{{1\over4}} \def\5{\overline }   \def\6{\partial } \def\7#1{{#1}\llap{/}}
\def\8#1{{\textstyle{#1}}}        \def\9#1{{\bf {#1}}}
\def\_#1{$\underline{\hbox{#1}}$} \def\^#1{$\overline{\hbox{#1}}$}

\def\<{\langle } \def\>{\rangle }  
 
\def \({\left( } \def \){\right) }

  \let\ex=\times   
      \let\and=\wedge
     
\def\|#1{{}_{\bigg|_{#1}}}

\ifx\Box  \else \fi

\def\pmbf#1{\setbox0=\hbox{${#1}$}   \kern-.025em\copy0\kern-\wd0
      \kern.05em\copy0\kern-\wd0     \kern-.025em\raise.0433em\box0 }
  \def\cc{{\cal C}} 
  \def\cg{{\cal G}} \def\ch{{\cal H}}
   
\def\cq{{\cal Q}}   
\def\inbar{\vrule height1.5ex width.4pt depth0pt}
\del
\def\ifundefined#1{\expandafter\ifx\csname#1\endcsname\relax}
\makeatletter \ifundefined{new@mathgroup} {} \else % \input{oldlfont.sty} \fi
\new@mathgroup\sffam
\new@internalmathalphabet\mathsf\sffam{cmss}{m}{n}
\def\psf{\fontfamily\sfdefault \fontseries\default@series
    \fontshape\default@shape\selectfont\mathsf}
\fi
\enddel
\def\ZZ{\relax{\sf Z\kern-.4em \sf Z}}  \def\IR{\relax{\rm I\kern-.18em R}}
\def\IN{\relax{\rm I\kern-.18em N}} \def\IP{\relax{\rm I\kern-.18em P}}
\def\IQ{\relax\,\hbox{$\inbar\kern-.3em{\rm Q}$}}
\def\IC{\hbox{\,$\inbar\kern-.3em{\rm C}$}}
\def\citen#1{\if@filesw \immediate\write \@auxout {\string\citation{#1}}\fi%
\@tempcntb\m@ne \let\@h@ld\relax \def\@citea{}%
\@for \@citeb:=#1\do {\@ifundefined {b@\@citeb}%
    {\@h@ld\@citea\@tempcntb\m@ne{\bf ?}%
    \@warning {Citation `\@citeb ' on page \thepage \space undefined}}%
    {\@tempcnta\@tempcntb \advance\@tempcnta\@ne
    \setbox\z@\hbox\bgroup\ifcat0\csname b@\@citeb \endcsname \relax
       \egroup \@tempcntb\number\csname b@\@citeb \endcsname \relax
       \else \egroup \@tempcntb\m@ne \fi \ifnum\@tempcnta=\@tempcntb
       \ifx\@h@ld\relax \edef \@h@ld{\@citea\csname b@\@citeb\endcsname}%
       \else \edef\@h@ld{\hbox{--}\penalty\@highpenalty
	      \csname b@\@citeb\endcsname}\fi
    \else \@h@ld\@citea\csname b@\@citeb \endcsname \let\@h@ld\relax \fi}%
 \def\@citea{,\penalty\@highpenalty\hskip.13em plus.13em minus.13em}}\@h@ld}
\def\@citex[#1]#2{\@cite{\citen{#2}}{#1}}%
\def\@cite#1#2{\leavevmode\unskip
  \ifnum\lastpenalty=\z@\penalty\@highpenalty\fi% highpenalty before
  \ [{\multiply\@highpenalty 3 #1%              % triple-highpenalties within.
  \if@tempswa,\penalty\@highpenalty\ #2\fi}]}   % and before note.
\makeatother % END OF   "CITE.STY"   (\citen can be used for citation numbers)

\def\beq{\begin{equation}} \def\eeq{\end{equation}} \def\eql#1{\label{#1}\eeq}
\def\bea{\begin{eqnarray}} \def\eea{\end{eqnarray}} 
\def\fnote#1#2{\begingroup\def\thefootnote{#1}\footnote{#2}
	   \addtocounter{footnote}{-1}\endgroup}

\def\plb#1 #2 {Phys. Lett. {\bf B#1} #2 }
\def\phr#1 #2 {Phys. Rep. {\bf  #1} #2 } 
\def\npb#1 #2 {Nucl. Phys. {\bf B#1} #2 }
\def\aph#1 #2 {Ann. Phys. {\bf #1} #2 }  
\def\jmp#1 #2 {J. Math. Phys. {\bf #1} #2 }
\def\prd#1 #2 {Phys. Rev. {\bf D#1} #2 }
\def\prl#1 #2 {Phys. Rev. Lett. {\bf #1} #2 }
\def\rmp#1 #2 {Rev. Mod. Phys.  {\bf #1} #2 }
\def\zpc#1 #2 {Z. Phys. {\bf #1C} #2 }
\def\cmp#1 #2 {Comm. Math. Phys. {\bf #1} #2 }
\def\mpl#1 #2 {Mod. Phys. Lett. {\bf A#1} #2 }
\def\ijmp#1 #2 {Int. J. Mod. Phys. {\bf A#1} #2 }

 \def\[{\left[} \def\]{\right]} 

\def\tbf#1:{{\noindent\bf #1:}}            \long\def\new#1\endnew{{\bf #1}}

\let\Ph=X  \let\Ps=Y
\def\eb{\bar\e}

\def\figuresonly{\pagestyle{empty}\figa\ve\figb\ve\figc\end{document}}
\long\def\old#1\endold{{\small #1}}         \def\oldansw{o } \def\cutansw{c }
\newcount{\npre} \npre=1  \def\negansw{s } \def\figansw{f } \def\textansw{t }
\def\ifpre{\ifnum\npre=1 } \def\ifsub{\ifnum\npre=0 }        \def\cut#1{#1}
\def\askversion{\message{
Preprint (p) / submit (s) / text only (t) / figures only (f):  (p/s/t/f)? }
    \read-1 to\answ \ifx\answ\negansw \npre=0 \else \npre=1 \fi
    \ifx\answ\figansw { } \else \def\figuresonly{ }   \fi
    \ifx\answ\oldansw \def\old##1\endold{{\small ##1}}\fi
    \ifx\answ\textansw \npre=2 \else \message{
Cut figures (use 'c' in case of memory problem):  (c/n)? }
    \read-1 to\answ\ifx\answ\cutansw \def\cut##1{}\npre=7\fi\fi \figuresonly }

\def\bpic{\begin{picture}} \def\epic{\end{picture}} \thicklines
\def\lab#1)#2#3{\put#1){\makebox(0,0)[#2]{\small #3}}}
\def\putlin#1,#2,#3,#4,#5){\put#1,#2){\line(#3,#4){#5}}} %\putlin(x,y,dx,dy,l)
\def\putvec#1,#2,#3,#4,#5){\put#1,#2){\vector(#3,#4){#5}}}

\newcount{\vmul} \newcount{\hdiv} % hor.length /=(\hdiv) vert.length *=(\vdiv)
\newcount{\wi}   \newcount{\he}   % width (height) of figure in \unitlength
\newcount{\hq}   \newcount{\vq}   % hor. (vert.) quantities
\newcount{\hoff} \newcount{\auxc} \newcounter{figco}   \def\npt{\circle*{2}}

\def\vlline{\put(-3,0){\line(1,0)6}}      % linemark for labels on vert. axis

\def\putvm#1{\mbox{\bpic(0,0)\funit=1pt\vlline\epic}}   % linemark (no label)

\def\hlabo{\put(0,0){\mbox{\bpic(0,0)\funit=1pt\put(0,-3){\line(0,1)3}\epic}}}

\def\Vpt#1,#2){\hq=#2\advance\hq by -#1 \multiply\hq by 2 \divide\hq by\hdiv
	       \vq=#1\advance\vq by #2 \multiply\vq by\vmul\put(\hq,\vq){\npt}}
\def\Vplo#1{\vbox{\hdiv=2\vmul=1 \figsca \auxc=\he \multiply\auxc by\vmul
    \hoff=\wi\divide\hoff by2 \stepcounter{figco}\message{[Fig. \arabic{figco}}
    \begin{center}\let\.=\Vpt \bpic(\wi,\auxc)(-\hoff,0) \figlab #1 \hlabo
    \put(-\hoff,0){\framebox(\wi,\auxc){}} \epic \\[5mm]
    Fig. \arabic{figco}: \figcap \end{center}} \vfil \message{]}}
\def\figsca{\unitlength=1.1pt \wi=500 \he=400} \let\funit=\unitlength
\begin{document}
\def\hannover{ITP--UH--03/95} \def\wien{TUW--95/01}
{\hfill hep-th/9505120   \vskip-5pt \hfill\wien \vskip-9pt\hfill\hannover}
\vskip 15mm \centerline{\hss\Large\bf
                        Orbifolds with Discrete Torsion  \hss}\ssk
	\centerline{\hss\Large\bf    and Mirror Symmetry    \hss}
\begin{center} \vskip 8mm
       Maximilian KREUZER\fnote{*}{e-mail: kreuzer@tph16.tuwien.ac.at}
\vskip 3mm
       Institut f"ur Theoretische Physik, Technische Universit"at Wien\\
       Wiedner Hauptstra\3e 8--10, A-1040 Wien, AUSTRIA
\vskip 6mm               and
\vskip 3mm
       Harald SKARKE\fnote{\#}{e-mail: skarke@itp.uni-hannover.de}
\vskip 3mm
       Institut f"ur Theoretische Physik, Universit"at Hannover\\
       Appelstra\3e 2, D--30167 Hannover, GERMANY

\vfil                        {\bf ABSTRACT}                \end{center}

For a large class of $N=2$ SCFTs, which includes minimal models and many
$\s$ models on Calabi-Yau manifolds, the mirror theory can be obtained as an
orbifold. We show that in such a situation the construction of the mirror
can be extended to the presence of discrete torsions.
In the case of the $\ZZ_2\ex\ZZ_2$ torus orbifold, discrete torsion between
the two generators directly provides the mirror model.
Working at the Gepner point it is, however, possible to understand this
mirror pair as a special case of the Berglund--H"ubsch construction.
This seems to indicate that the $\ZZ_2\ex\ZZ_2$ example is a mere coincidence,
due to special properties of $\ZZ_2$ twists, rather than a hint at
%% an indication of
a new mechanism for mirror symmetry.

\vfil %\noindent \hannover\\[1pt] \wien\\[3pt] April 1995 \msk
\thispagestyle{empty} \newpage
\pagestyle{plain} % \setcounter{page}{1}
\ifsub \baselineskip=20pt \else \baselineskip=14pt \fi

\def\Z#1{{\ZZ_{#1}}}   \def\pha#1,#2.{\ph_{#1}^{(#2)} } \def\sig#1{(-1)^{#1}}
\def\^#1{\overline{#1}} \def\A{{\5\a}} \def\X{{\^X}} \def\W{{\^W}}

\section{Introduction}

In $N=2$ superconformal field theories  mirror symmetry (MS) is nothing
but a change of sign of the left-moving $U(1)$ charge.
In some cases, like the minimal models, the $U(1)$ charge conjugation matrix
is a simple current modular invariant, i.e. a modular invariant that
relates only primary fields on the same orbits of the simple currents under
fusion (simple currents, by definition, have a unique fusion product,
so that all primary fields are organized into orbits; they can be shown
to imply discrete symmetries of the conformal field theory \cite{sc90r}).
Via the correspondence between simple current modular invariants and orbifolds
\cite{sci} this implies that the mirror theory of any orbifold of such a model
is also an orbifold of the same conformal field theory.\footnote{
	Note that a general $N=2$ theory is a product of some neutral CFT with
        a $U(1)$ current algebra. This product theory, however, does not have
	a product modular invariant. Hence, although charge conjugation
	in the $U(1)$ factor is always described by a simple current modular
	invariant, this is not true in general for the complete $N=2$ theory.}
For the case of minimal models this was first shown by Greene and
Plesser~\cite{gp}. If the conformal field theory is related to a $\s$
model on a Calabi-Yau manifold, then this simple observation has very
non-trivial consequences \cite{co}.

More generally, since MS exchanges the (c,c) ring and the (a,c) ring
\cite{lvw}, we have a chance to obtain the mirror model of a diagonal
SCFT by orbifolding only if the symmetry group is large enough to project
out all fields in the chiral ring of the diagonal theory.\footnote{
	Throughout this paper, we work with the internal SCFT
	before the generalized GSO projection. In order to make contact with
	related $\s$ models we first need to project to
	integral charges. This just amounts to a restriction on the allowed
	twist groups and discrete torsions, so that these cases are covered
	by our more general discussion.}
For Landau--Ginzburg (LG) models \cite{mvw,lvw} this is the case if a
non-degenerate superpotential is a sum of $N$ monomials in $N$ superfields.
Indeed, Berglund and H"ubsch (BH) \cite{bh} found a map among orbifolds that
produces the mirror theory for Landau--Ginzburg models and for the Calabi--Yau
hypersurfaces based on this type of polynomials. More recently, it was checked
by two different approaches that this map indeed produces the mirror theory:
In \cite{eg} the elliptic genera of the dual orbifolds were compared. In
\cite{mmi} the map sending a monomial representation of a chiral ring element
to a twist group element was constructed explicitly, and the twist selection
rules of the orbifold were shown to be consistent with the OPE selection rule
of the original theory.

In the present paper we show how this construction
can be extended to arbitrary orbifolds, including discrete torsions \cite{dt}.
It is well known that the BH construction cannot be extended to more general
LG models \cite{nms,aas,lgt};
in that case a construction of the mirror is available only in the
geometrical framework of toric varieties \cite{toric}. For these,
however, it is not yet clear how the freedom of choosing discrete torsions in
the CFT approach can be realized.

Torus orbifolds provide a much smaller class of internal conformal field
theories. The corresponding numbers of generations and anti-generations
were listed in \cite{fiq}.
This list features
two mirror pairs of spectra, each of them requiring discrete torsion (DT).
This led to some speculations that DT might be essential for
mirror symmetry \cite{privat}. More recently, Vafa and Witten
\cite{vw} reinvestigated the $\ZZ_2\ex\ZZ_2$ orbifold, whose striking
feature is that the inclusion of discrete torsion directly provides the mirror
model.
%% This mechanism
The situation is different, however,
%% cannot be applied to
for the second mirror pair
which requires DT at both sides of the correspondence \cite{fiq,vw}.
We will show that the $\ZZ_2\ex\ZZ_2$ case can be understood as a special
case of the BH construction if we reinterpret the DT in a way that is
specific to $\ZZ_2$ twists.
In contrast to this, the $\ZZ_3\ex\ZZ_3$ and the $\ZZ_6\ex\ZZ_6$ orbifolds
with $\ZZ_3$ torsions, despite having the same non-singlet spectrum,
have different
numbers of singlets and gauge bosons. Consequently, there is no generalized
BH map between these two models, although a mirror model is straightforward
to construct in each case (the calculation of the singlet number is, in fact,
less tedious for the mirror models).
This is consistent with
the general picture that mirror symmetry, if at all related to orbifolding,
operates within the subclass of symmetric orbifolds, but can be extended to
accomodate discrete torsions.
% Eastwood -> singlets of 3-generation model ...

In section 2 we will discuss the modding of quantum symmetries, i.e.
symmetries that follow from the twist selection rules for orbifolds,
and show how this technique can be used to explicitly construct the mirror
twists and torsions whenever the mirror theory
of some diagonal $N=2$ SCFT can be obtained as an abelian orbifold of some --
possibly different -- diagonal theory.
In the appendix we demonstrate this mechanism for an asymmetric
$(\ZZ_3)^5$ orbifold in the tensor product $1^9$ of minimal models that has
no generations and no anti-generations at all.
Here the rank of the twist group for the mirror model is 6, rather than 4,
as would be the case without~DT.
In section 3 we discuss the two candidate mirror pairs of torus orbifolds by
applying the methods of section 2 at Gepner points in moduli space.

\section{Mirror map and discrete torsion}

Consider a situation where the mirror model to a diagonal $N=2$ theory $\cc$
is given by an abelian orbifold of a possibly different diagonal $N=2$ theory
$\5\cc$. In case of the minimal models we have $\cc=\5\cc$ \cite{gp},
but for Landau--Ginzburg
models that are described by sums of polynomials of the type
\beq
     W_{\rm loop}=X_1^{a_1}X_2+\ldots+X_{n-1}^{a_{n-1}}X_n+X_n^{a_n}X_1,
\eeq
\beq
     W_{\rm chain}=X_1^{a_1}X_2+\ldots+X_{n-1}^{a_{n-1}}X_n+X_n^{a_n},
\eeq
it was first observed in \cite{bh} that one has to twist the theories
described by the respective `transposed' potentials
\beq
     \W_{\rm loop}=\X_1\X_2^{a_1}+\ldots+\X_{n-1}\X_n^{a_{n-1}}+\X_n\X_1^{a_n},
\eeq
\beq
     \W_{\rm chain}=\X_1^{a_1}+\X_1\X_2^{a_2}+\ldots+\X_{n-1}\X_n^{a_n}
\eeq
in order to obtain the mirror models.
(The minimal models with diagonal modular invariant correspond to Fermat
potentials $X^a$, i.e. to the case $n=1$, whereas the $E_7$ invariant and
the $D$ series can be described by $W_{\rm chain}$ with $n=2$.)

Obviously, the SCFT based on $\W$ is in general different from the one based
on $W$.
The orders of the maximal phase symmetry groups
of $W$ and $\W$, however, are equal. Explicitly,
\beq
     |\cg_{\rm loop}|=a_1a_2\ldots a_n-\sig n, \hspace{45pt}
     |\cg_{\rm chain}|=a_1a_2\ldots a_n.
\eeq
$\cg$ is generated by
\beq
     \r_lX_j=e^{2\p i\,\ph^{(l)}_j} X_j, ~~~~~~ \ph^{(l)}_{l+j}={\sig{n-j}
     a_l\ldots a_{l+j-1}\0|\cg_{\rm loop}|} ~~~ \hbox{for} ~~ 0\le j<n
\eeq
for loop potentials and by
\beq
     \r_lX_j=e^{2\p i\,\ph^{(l)}_j} X_j, ~~~~~~~~
     \pha j,l.={\sig{l-j+1}\0a_j\ldots a_l} ~~~ \hbox{ for }1\le j\le l,~~~~
     \pha j,l.=0 ~~~ \hbox{ for }j>l
\eeq
for chain potentials.

The transformation of $X_1$
determines the complete group action, hence $\cg$ is cyclic. % and it
$\cg$ is generated by $\r_n$ in both cases;
for loop potentials any transformation $\r_i$
generates the whole group. It is, nevertheless, useful to consider the
complete set of transformations $\r_i$, because it can be checked that the
phase of the determinant of $\r_i$ is equal to $2\p$ times the charge of the
field $\X_i$.
Furthermore, the phase of $\det \r$ essentially determines the
%% charge of the ground state
left-right asymmetry of the U(1) charges of all states
in the sector  twisted by $\r$ in the orbifold
theory \cite{viv} (for details see \cite{mmi}).

This suggests that the mirror map should map the chiral
field $\X_i$ to a field in the sector twisted by $\r_i$,
and consequently,
because of the chiral ring structure and the twist selection rules, a monomial
$\prod\X_i^{\A_i}$ into the sector twisted by~$\prod \r_i^{\A_i}$.
The number of states in such a sector and their charges have been
shown to be consistent with the identification of the two
theories as a mirror pair \cite{mmi}.

Further evidence for the identification comes from a consideration of discrete
symmetries of the SCFT defined by the superpotential $W$:
Since the original LG model is symmetric, all moduli come from
the (c,c) ring. But the maximal phase symmetry projects out all (c,c) states
(as is necessary for obtaining a SCFT with asymmetric charges for the
orbifold).
Assume that the orbifold $W/\cg$, which has the correct charge
degeneracies of the chiral ring and a set of OPE selection rules that is in
one-to-one correspondence with the vanishing relations of the chiral ring,
lies in the moduli space of the mirror SCFT of $\W$. Then we can conclude that
it must indeed be the mirror model since all moduli are fixed by the presence
of the quantum symmetry that arises from the modding by $\cg$.
Furthermore, the elliptic genera of the two theories have been shown
to coincide \cite{eg}. That information goes beyond the chiral ring, but it
is insensitive to the SUSY preserving moduli of the conformal field theory.

Returning to the general discussion, we consider a ring basis $\X_i$ of
chiral fields of some diagonal superconformal theory $\5\cc$, which we assume
to be the mirror model to an abelian orbifold $\cc/\cg$ of another diagonal
theory $\cc$.
Then the mirror map must induce a map $\X_i\to\r_i$, where $\r_i\in\cg$ is
the twist of the sector that contains the image of $\X_i$
under the mirror map.
Since $\cc/\cg$ and $\5\cc$ are isomorphic CFTs (up to a redefinition of
$U(1)$ charges), any orbifold of one
of the theories must have a counterpart as an orbifold of the other theory.
In particular, the mirror model of $\cc$ must be described by the orbifold
$\5\cc/\5\cg$ with a certain symmetry group $\5\cg$  of $\5\cc$ that is
isomorphic to $\cg$, because we can obtain $\cc$ as
an orbifold of $\cc/\cg$ with respect to the quantum symmetry that comes with
the twist group $\cg$ (see below).

Our aim is to give an explicit description of the mirror orbifold in case of a
general quotient $\5\cc/\5\ch$ with respect to a subgroup
$\5\ch\subseteq\5\cg$ of the symmetry group $\5\cg$ of $\5\cc$, where we allow
arbitrary discrete torsions (as a concrete
example, we can keep in mind the case of the BH construction).
If some group element $\5 g\in\5\ch$ acts on a diagonal basis $\5 X_i$
of the chiral ring of $\5\cc$ as
$\5 g \5 X_i=\5 \g_i \5 X_i$ (with phases $\5 \g_i$), then it acts on the
chiral primary states (which are all in the untwisted sector) as
\beq \5 g\(\prod\X_i^{\A_i}|0\>\)=
         \(\prod\5\g_i^{\A_i}\)\(\prod\X_i^{\A_i}|0\>\).   \eeq
Then the mirror twist $q_{\5 g}$, % of this action,
which is a symmetry transformation of $\cc/\cg$, has to satisfy
\beq
     q_{\5 g}\((\ldots)|\prod\r_i^{\A_i}\>\)=
        \(\prod\5\g_i^{\A_i}\)\((\ldots)|\prod\r_i^{\A_i}\>\),
\eeq
where $|\r\>$ is the ground state of the sector that is twisted by
$\r=\prod\r_i^{\A_i}$.
This can be achieved with the following rule:
To any classical symmetry transformation $\bar g\in\5\ch$ of $\5\cc$ we
assign a quantum symmetry transformation \cite{q}
$q_{\bar g}$ of $\cc/\cg$, which acts trivially on all fields in $\cc$
% (thus ensuring $\det {q_{\bar g}}_{|_\r}= \det q_{\bar g}=1$),
and has discrete torsions with elements $\r_i$ of $\cg$
that are equal to the action of $\bar g$ on $\X_i$,
i.e. $\e(q_{\bar g}, \r_i) =\5\g_i$
(remember that a DT $\e(g,h)$ between g and h implies that $g|h\>$ picks up an
extra factor of $\e(g,h)$ compared to the case without DT \cite{dt}).
Torsions between elements $\bar g,\bar h\in\5\ch$ are translated
to torsions between $q_{\bar g}$ and $q_{\bar h}$.
% , and $K_{q_{\bar g}}=K_{\bar g}$.
Modding by a subgroup $\5{\cal H}$ of the full group $\5\cg$ of % phase
symmetries of $\5\cc$ hence corresponds to modding $\cc$ by $\cg\otimes
\cq_\ch$, where $\cq_\ch$ is the group of quantum symmetry transformations
$q_{\bar g}$ that correspond to the transformations $\5g\in\5\ch$.
The $q_{\bar g}$ act trivially on $\cc$ and have the described discrete
torsions with $\r_i\in\cg$ and among themselves.

Since the elements of $\cq_\ch$ are trivial symmetries of $\cc$
-- they act on $\cc$ like the identity --
we expect that it should be possible to get rid of them and to replace the
modding with twist group $\cg\otimes\cq_\ch$ by a modding with an equivalent
twist group $\ch\subseteq\cg$ that does not contain any `pure' quantum
symmetries. Indeed, this is possible by repeated application of the following
lemma.

{\it Reduction of quantum symmetries:}
If an abelian twist group has a quantum generator $q$ of order $n$,
we can rewrite the twists in such a form that $q$ has non-trivial torsion
$\e(q,g)=\z_n=e^{2\p i/n}$ with only a single generator $g$ whose order $N$
is a multiple of $n$. Then the original orbifold is
equivalent to the one with $q$ omitted and with $g$ replaced by $g'=g^n$.
%$g'=g^{`n}$.

In order to prove this statement we have to compare the conformal field
theories $\cc_0=\cc/(\ch\otimes\cg')$ and
$\cc_q=\cc/(\ch\otimes\cg\otimes\cq)$, where $g'$, $g$ and $q$ generate
the cyclic groups $\cg'$, $\cg$ and $\cq$, respectively
(the discrete torsions are of the form described above; $\ch$ could
be non-abelian without spoiling the argument). The latter
orbifold is a quotient by a quantum symmetry:
$\cc_q=\cc_g/\cq$ with $\cc_g=\cc/(\ch\otimes\cg)$. It is obvious that this
quotient eliminates all sectors with a twist by $g^l$ whenever $l$ is not
a multiple of $n$.
More explicitly, the $\cq$ projection keeps exactly
the $g$-invariant states in $\cc_0$. Since all states $|i\>$ in $\cc_0$
are invariant under $g'$, the transformation of $|i\>$ under $g$ is given
by a phase $(\z_n)^{a_i}$ in a diagonal basis of the Hilbert space.
Since $q$ is a pure quantum symmetry, a twist by $q^a$ does not change
any quantum number of a state except for its transformation property under $g$.
Hence there is exactly one copy of each state in $\cc_0$, namely the one in
the sector twisted by $q^{a_i}$, which survives the $g-$projection in the
orbifold $\cc_q=(\cc/(\ch\otimes\cg))/\cq$.

If there are no discrete torsions among the generators of the quantum subgroup
$\cq$ of a twist group $\cg=\cq\otimes\ch$,
a strict separation between pure classical and pure quantum
symmetries is maintained throughout this elimination process.
Each elimination of a quantum generator simply amounts to reducing
the group of classical symmetries to the subgroup that has trivial
torsion with this generator. So we end up with the group of symmetries
that have vanishing torsion with all the elements of $\cq$.

Returning to the case of an %% In the case of
orbifold construction of the mirror models we have to twist
$\cc$ by the group $\cg\otimes \cq_\ch$, where the modding by $\cg$
produces the mirror SCFT of $\5\cc$ and $\cq_\ch$ corresponds to the modding
$\5\ch$ of $\5\cc$ for which we want to construct the mirror model.
If there are no discrete torsions in $\5\ch$,
% Thus we get
elimination of the quantum twists thus gives us
the `dual' subgroup $\cal H$ satisfying
$|{\cal H}||\5{\cal H}|=|\cg|=|\5\cg|$ without any torsions, as in
the original work of Berglund and H\"ubsch\cite{bh}.
In the case of torsions the situation is more complicated and torsions
among the quantum symmetry generators will induce torsions among the
classical twists. In particular,  the order of the group $\ch$ we
end up with can be larger than $|\cg|/|\5\ch|$. In the appendix this is
demonstrated for an $\5\ch=(\ZZ_3)^5$ orbifold of the model $1^9$ for which
$\cg=(\ZZ_3)^9$. Without torsions, the dual orbifold comes from a twist
group of rank 4, but in the example that we discuss we end up with
$\ch=(\ZZ_3)^6$.

\section{Torus orbifolds}

The mirror pair considered in ref. \cite{vw} consists of the
$\ZZ_2\ex\ZZ_2$ orbifold of the product of three tori without and with
DT between the two $\Z2$ factors.
The twist group is generated by $\s_1\s_2$ and $\s_1\s_3$, where
$\s_i:z_i\to-z_i$ changes the sign of the complex coordinate of the $i^{th}$
torus.
In the Weierstra\3 normal form of the embedding of an elliptic curve in
$\IP^2$,
\beq x_0x_2^2=4x_1^3-g_2x_1x_0^2-g_3x_0^3 \eql{wei}
this symmetry corresponds to $x_2\leftrightarrow -x_2$.
At $g_2=0$, which corresponds to a torus with complex structure
modulus $\t=\exp(2\p i/3)$,
we have an additional $\Z3$ symmetry which will allow us to
apply the BH construction.
To this end consider the potential
\beq
     W_D=X^2Y+Y^3+Z^3 ,
\eql{wd}
obtained from (\ref{wei}) with $g_3=g_2=0$ by a change of variables,
and its phase symmetries
\beq
     \r_X(X,Y,Z)=(-X,Y,Z), ~~~ \r_Y(X,Y,Z)=(\z_6X,\z_3^2Y,Z), ~~~
     \r_Z(X,Y,Z)=(X,Y,\z_3^2Z).
\eeq
Since $\r_X=\r_Y^3$ the group of phase symmetries is generated by $\r_Y$ and
$\r_Z$; a torus is obtained by modding the $\Z3$ generated by
$j=\r_Y^2\r_Z^2$.
%At a general value of the
The first two terms in $W_D$ are just the LG representation of the  $D$
invariant of the minimal model at level 4 \cite{mvw}, which %. This model
is the $\ZZ_2$ orbifold of the $A$ invariant. % of the same minimal model;

The classical $\ZZ_2$ symmetry $\r_X$ of $W_D$ corresponds
to the quantum $\ZZ_2$ symmetry of
\beq
     W_A=X^2+Y^6+Z^3
\eql{wa}
modded by its canonical $\ZZ_6$ symmetry
\beq j(X,Y,Z)=(\z_2X,\z_6Y,\z_3Z).\eeq
Therefore the product of three tori, modded by $\ZZ_2\ex\ZZ_2$, can be
represented by
\beq
     W=\sum_{i=1}^3\(X_i^2+Y_i^6+Z_i^3\),
\eeq
modded by $(\ZZ_6)^3\ex(\IQ\ZZ_2)^2$.
Let us choose generators $j_i^{(2)}=j_i^3$ (of order 2) and $j_i^{(3)}=j_i^2$
(of order 3)
for each of the $\ZZ_6=\ZZ_2\ex\ZZ_3$ factors and
$q_{12}$ and $q_{13}$ for $\IQ\ZZ_2$, such that $q_{12}$ has torsion $-1$
with $j_1^{(2)}$ and $j_2^{(2)}$, $q_{13}$ has torsion $-1$ with $j_1^{(2)}$
and $j_3^{(2)}$ and all other torsions between quantum and classical
symmetries are trivial.
The case without torsion between the two $\ZZ_2$
factors acting on the torus corresponds to trivial torsion between $q_{12}$
and $q_{13}$.
In this case the rules given above can easily be used to
show that the modding of $W$ by $(\ZZ_6)^3\ex(\IQ\ZZ_2)^2$ can be
reduced to a modding by $(\ZZ_3)^3\ex\ZZ_2$ generated by $j_i^{(3)}, i=1,2,3$
and $j^{(2)}=j_1^{(2)}j_2^{(2)}j_3^{(2)}$.

The mirror of this model is $W$ modded by the group
$(\ZZ_6)^6\ex(\IQ\ZZ_3)^3\ex\IQ\ZZ_2$
generated by
\beq \r_{X_i},\qd \r_{Y_j},\qd \r_{Z_k},\qd q_{j^{(3)}_l},\qd q_{j^{(2)}}. \eeq
The cancellations of the quantum generators take place independently
among generators of order 2 and generators of order 3:
We can replace
\beq \r_{X_i},\qd \r_{Y_k}^{(2)},\qd q_{j^{(2)}}\eeq
by
\beq j_i^{(2)},\qd \r_{X_1}\r_{X_2}, \qd \r_{X_1}\r_{X_3},\eeq
and we can replace each of the sets
\beq \r_{Y_i}^{(3)},\qd \r_{Z_i},\qd q_{j^{(3)}_i}\qd  \eeq
(with $i=1,2,3$) by ${j^{(3)}_i}$.
So the mirror model is given by $(\ZZ_6)^3\ex(\ZZ_2)^2$, where the
$\ZZ_6$ factors again correspond to the canonical $\ZZ_6$ symmetries
of the three parts and the two $\ZZ_2$'s are generated by $\r_{X_1}\r_{X_2}$
and $\r_{X_1}\r_{X_3}$.

As the generators $\r_{X_i}$
act only on the trivial fields $X_i$ which do not contribute
to the chiral ring, their action depends only on the twist of chiral state in
the orbifold.
But a symmetry
acting only on the twisted vacua, and not on the chiral fields of the
untwisted theory, may as well be regarded as a quantum
symmetry with properly chosen discrete torsion.\footnote{
The fact that $\Z2$ symmetries acting on trivial fields can mimic DT
has already been exploited in \cite{aas}.}
To see how this works in detail
we use the result of Intriligator and Vafa \cite{viv} for the action of a
group element $g$
on the ground state $|h\>$ of a twisted sector\footnote{
	Our conventions differ from \cite{viv} by a factor $(-1)^{K_gK_h}$ in
	the definition of the discrete torsion $\e(g,h)$, where $(-1)^{K_g}$
	is the discrete torsion between a group element $g$ and the $\ZZ_2$
	twist that generates the Ramond sector.}
\beq
     g|h\>=\e(g,h)\,{\det}_h(g) ~|h\>,
\eeq
where $\det_h(g)$ is the determinant of $g$ restricted to the subspace of
superfields of the LG theory that are twisted by $h$
(we work in a basis where both $g$ and $h$ are diagonal).
If $g$ acts only on trivial fields, then
$\det_h(g)=\det_g(h)=(-1)^{n_{g,h}}$,
where $n_{g,h}$ is the number of trivial fields on which the actions of $g$
and $h$ are $-1$. Hence we can replace $g$ by a symmetry that acts trivially
on all fields (i.e. a quantum symmetry) and, at the same time, $\e(g,h)$ by
$(-1)^{n_{g,h}}\e(g,h)$. It is easily checked %% with the above formula
that this substitution produces the correct transformations also in
all twisted sectors.
In our above representation of the mirror of the $\ZZ_2\ex\ZZ_2$ orbifold we
can thus
identify $\r_{X_1}\r_{X_2}$ with $q_{12}$ and $\r_{X_1}\r_{X_3}$
with $q_{13}$, with the same torsions between classical and quantum
generators as above, but with nontrivial DT of $-1$ between $q_{12}$ and
$q_{13}$.
Thus we have indeed related the $\ZZ_2\ex\ZZ_2$ torus orbifold without
torsion to the $\ZZ_2\ex\ZZ_2$ torus orbifold with torsion.

In the list of torus orbifolds in ref. \cite{fiq} there is another
candidate for a mirror pair: The $\ZZ_3\ex\ZZ_3$ orbifold with
the actions of the generators $g_i$ on the complex coordinates $z_i$
given by multiplication with $(\z_3,1,\z_3^{-1})$ and
$(1,\z_3,\z_3^{-1})$, and with a discrete torsion of $\e(g_1,g_2)=\z_3$ among
the generators, has the spectrum $n_{\5{27}}=27$ and $n_{27}=3$.
The $\ZZ_6\ex\ZZ_6$ orbifold with
the actions of the generators $g_i'$ given by $(\z_6,1,\z_6^{-1})$ and
$(1,\z_6,\z_6^{-1})$ and with the same discrete torsion $\e(g'_1,g'_2)=\z_3$
has $n_{\5{27}}=3$ and $n_{27}=27$.
We will show now, however, that this mirror pairing of spectra cannot be
understood by the BH construction. Moreover, if we compute the complete
massless spectrum, it turns out that the numbers of gauge bosons disagree.

In order to implement the $\ZZ_3\ex\ZZ_3$ symmetry in a LG model
we have to represent each of the three tori by a potential $W_i,
i=1,2,3$ %that is either
like, for example, $W_A$ or $W_D$ (cf. eqs. (\ref{wa},\ref{wd})).
\del
\beq
     \5W_D=X^2+Y^3X+Z^3,
\eql{wdb}
or
\beq
     W_{(1)}=X^3+Y^3+Z^3.
\eeq
{\bf !!! es gibt mehr M"oglichkeiten, wie z.B.
	 $X^3+ZY^2+YZ^2$ oder $X^4+XY^3+Z^2$ !!! }\\
\enddel
The $\ZZ_3\ex\ZZ_3$ orbifold is then represented by the LG potential
$W=W_1+W_2+W_3$ modded by the group $\cal H$ generated by
$j_1$, $j_2$, $j_3$, $g_{13}^{(3)}$, $g_{23}^{(3)}$, with nontrivial DT only
between $g_{13}^{(3)}$ and $g_{23}^{(3)}$.
The latter
generators (of orders~3) correspond to the generators acting on the complex
coordinates $z_i$ by multiplication with $(\z_3,1,\z_3^{-1})$ and
$(1,\z_3,\z_3^{-1})$.
%The set of canonical generators can be replaced by
\del
$j_1^{(2)}$, $j_2^{(2)}$,
$j_3^{(2)}$ (of orders 2) and $j_1^{(3)}$, $j_2^{(3)}$, $j_3^{(3)}$,
$g_{13}^{(3)}$, $g_{23}^{(3)}$ (of orders~3),
\\{\bf !!! zu $g_{13}^{(3)}$, $g_{23}^{(3)}$ sollte man auch noch ein paar
	 Worte sagen (zugegeben, wenn man lange genug dar"uber nachdenkt,
	 findet man schon raus, was damit gemeint sein mu\3 ...) !!! }\\
with nontrivial DT only
between $g_{13}^{(3)}$ and $g_{23}^{(3)}$.
\enddel
The BH mirror of this model is described by $\5W=\5W_1+\5W_2+\5W_3$
%(with $\5W_A=W_A$ and $\5W_{(1)}=W_{(1)}$)
modded by some group $\5{\cal H}$.
%generated again by elements of orders 2 and 3.
Here we note again that the
construction of the mirror twist group takes place independently for
each prime number.
In particular going from $\cal H$ to a different twist group
$\tilde{\cal H}$ by dropping $g_{13}^{(3)}$ and $g_{23}^{(3)}$ does
not affect the mirror construction in the rank 2 sector.
But $W/\tilde{\cal H}$ is just the product of three tori, whose mirror
$W/\5{\tilde{\cal H}}$ is well known to be the product of three tori, again.
If we could identify $W/\5{\cal H}$ with a $\Z6\ex\Z6$ torus orbifold,
we could identify $W/\5{\tilde{\cal H}}$ with a $\Z2\ex\Z2$ torus orbifold,
which is clearly a contradiction.

The observation that there is no BH construction of a mirror map relating the
two torus orbifolds under consideration suggests that, after all, they
may not be a real mirror pair. We should thus perform a more stringent
test of this possibility. Since we have a represention of the orbifolds as
exactly solvable conformal field theories, the natural next step is to compute
the complete massless spectrum. There exists an extensive list of such spectra
\cite{sy}, which should contain the information we need.
Indeed, we find the numbers $\{27,3\}$ for $\{n_{27},n_{\5{27}}\}$ among the
(2,2) vacua that were constructed in \cite{sy}
from the tensor products $1^9$ and $1^34^3$ of minimal models. For $1^9$,
which corresponds to $W_i=X^3+Y^3+Z^3$, the
only  result that is consistent with the tables supplement of \cite{sy}
is $n_S=252$ and $n_V=8$, where $n_S$ is the number of singlets and $n_V$
is the number of extra gauge bosons. For the product $1^34^3$, which
corresponds to $W_i=X^3+Y^6+Z^2$, there is also
a second possibility, which is $n_S=213$ and $n_V=5$.

The $\ZZ_3\ex\ZZ_3$ orbifold can be represented as a phase orbifold of
$1^9$, hence we know that its spectrum must be given by $n_S=252$ and $n_V=8$.
Unfortunately, this is not quite enough for our purpose since
we still need to
compute at least one of the number $n_S$ and $n_V$ for the $\ZZ_6\ex\ZZ_6$
orbifold.
Once more, the BH construction can be used to simplify the calculation:
Starting with the representation of the $\ZZ_6\ex\ZZ_6$ torus orbifold
as a $(\ZZ_6)^5$ quotient of the LG model with the
potential $W=W_1+W_2+W_3$ with $W_i=X_i^3+Y_i^6+Z_i^2$
it is straightforward to construct the BH mirror.
We find a $\ZZ_6\ex\ZZ_3\ex\ZZ_3$ quotient of $W$
with generators $g_1=j_1j_2j_3$, $g_2=(j_1)^2$,
$g_3=(j_2)^2$, where $j_i$ generates the canonical $\ZZ_6$ twist of $W_i$,
and the only torsion is $\e(g_2,g_3)=\z_3$.
For that orbifold the number of extra $U(1)$'s can easily be computed to be
$n_V=5$ (all of them come from descendents due to the $U(1)$'s of the factors
of the tensor products). Hence the two torus orbifolds indeed have different
singlet spectra and do not form a mirror pair.\footnote{
	As a check, the same calculation can be repeated for the
	$\ZZ_3\ex\ZZ_3$ orbifold, represented as a $(\ZZ_6)^3\ex(\ZZ_3)^2$
	quotient. Here the BH construction of the mirror results in the
	$(\ZZ_6)^3$ quotient with $g_1=j_1j_2j_3$, $g_2=j_1$,
	$g_3=j_2$ and $\e(g_2,g_3)=\z_3$. The number of extra gauge bosons
	turns out to be 8 (the 3 additional chiral states with charge 0 and
	conformal weight 1 can be represented as $6^{th}$ powers $J_s^{(i)}$
	of the spinor currents $J_s^{(i)}$ that generate spectral flow in the
	$c=3$ factors of the tensor product).}

\appendix\section{\hspace*{-21pt}ppendix}

As an example for the general construction of the mirror orbifold
consider the model $(1)^9$, which can
be described by the LG--potential
\beq W=X_1^3+X_2^3+X_3^3+Y_1^3+Y_2^3+Y_3^3+Z_1^3+Z_2^3+Z_3^3.\eeq
The symmetry group $\cg=(\ZZ_3)^9$ is generated by
\beq \r_{X_i},\qd \r_{Y_i},\qd \r_{Z_i},\qqd i=1,2,3,\eeq
%The orbifold of $W$ by t
where $\r_{X_i}$ acts on $X_i$ with a phase $1/3$ and leaves all other
fields invariant, and similarly for $Y$ and $Z$.
We choose the twist group $\5\ch$ that is generated by
\beq \!\!\!\!\!\!\!
    \r_X=\r_{X_1}\r_{X_2}\r_{X_3},\qd\!\! \r_Y=\r_{Y_1}\r_{Y_2}\r_{Y_3},
      \qd\!\! \r_Z=\r_{Z_1}\r_{Z_2}\r_{Z_3}, \qd\r_1=\r_{X_1}\r_{Y_1}\r_{Z_1},
    \qd\!\! \r_2=\r_{X_2}\r_{Y_2}\r_{Z_2} \!\!   \eeq
% (it contains the canonical twist $j=\r_X\r_Y\r_Z$). If we choose
and the discrete torsions %to be
\bea &\e(\r_X, \r_Y)=\e(\r_X, \r_Z)=\e(\r_Y, \r_Z)=1,&\cr
   &\e(\r_X,\r_i)=%\e(\r_X,\r_2)=
    \e(\r_Y,\r_i)=%\e(\r_Y,\r_2)=
    \e(\r_Z,\r_i)=%\e(\r_Z,\r_2)=&\cr&
    \e(\r_1,\r_2)=\exp(4\pi i/3).   &  \eea
In a matrix representation
{\small
\beq
     \hbox{\large$\5\ch\sim$} \pmatrix{ 1 & 1 & 1 & 0 & 0 & 0 & 0 & 0 & 0 \cr
     	0 & 0 & 0 & 1 & 1 & 1 & 0 & 0 & 0 \cr
	0 & 0 & 0 & 0 & 0 & 0 & 1 & 1 & 1 \cr
     1 & 0 & 0 & 1 & 0 & 0 & 1 & 0 & 0\cr0 & 1 & 0 & 0 & 1 & 0 & 0 & 1 & 0\cr}
	,~~~~~~
     \hbox{\large$\5\e\sim$} \pmatrix{ 0 & 0 & 0 & 2 & 2 \cr
	0 & 0 & 0 & 2 & 2 \cr 0 & 0 & 0 & 2 & 2 \cr
	1 & 1 & 1 & 0 & 2 \cr 1 & 1 & 1 & 1 & 0 \cr} , 		\label H
\eeq}%
where the phases of the twists and discrete torsions are all in units of $1/3$.
% we obtain a model whose
The `Hodge diamond' of this model can be calculated with the
methods of refs. \cite{viv}. In practice we used a computer program that
we developed for the investigation of ref. \cite{ade} to obtain
\beq \matrix{ &   &   & 1 &   &   &   \cr
              &   & 0 &   & 3 &   &   \cr
              & 0 &   & 0 &   & 3 &   \cr
            1 &   & 0 &   & 0 &   & 1 \cr
              & 3 &   & 0 &   & 0 &   \cr
              &   & 3 &   & 0 &   &   \cr
              &   &   & 1 &   &   &   \cr},  \quad\label{hd}
\eeq
%(this means that
i.e. the charge degeneracy is 3 for chiral ring elements with
left and right $U(1)$ charges
$(q_L,q_R)\in\{(1,0),(2,0),(1,3),(2,3)\}$, whereas there are no chiral
states with $q_R=1$ or $q_R=2$.
% More precisely, there are 3 chiral primary states with left and right
% $U(1)$ charges $(q_L,q_R)=(1,0)$ and no states with $(q_L,q_R)=(1,0)$;
% all other numbers should be clear from (\ref{hd}).

According to the construction we gave in section 2, the mirror to this
model is given by the quotient of $W$ by the group
$\cg\otimes \cq_\ch$ that is generated by
\beq
     \r_{X_i},\qd\r_{Y_i},\qd\r_{Z_i},\qd q_X,\qd q_Y,\qd q_Z,\qd q_1,\qd q_2.
\eeq
The $q$-generators act trivially on all fields, but have torsions
given by
\bea
     &\eb(\r_{X_i},q_X)=\eb(\r_{Y_i},q_Y)=\eb(\r_{Z_i},q_Z)=&\cr
     &\eb(\r_{X_1},q_1)=\eb(\r_{Y_1},q_1)=\eb(\r_{Z_1},q_1)=
     \eb(\r_{X_2},q_2)=\eb(\r_{Y_2},q_2)=\eb(\r_{Z_2},q_2)=&\cr
   &\eb(q_X,q_1)=\eb(q_X,q_2)=\eb(q_Y,q_1)=\eb(q_Y,q_2)=
    \eb(q_Z,q_1)=\eb(q_Z,q_2)=&\cr&\eb(q_1,q_2)=\exp(4\pi i/3),  &
\eea
with all other torsions being trivial.
Changing our set of generators to
\bea &\r_X,\qd\r_{X_1}/\r_{X_2},\qd\r_Y,\qd\r_{Y_1}/\r_{Y_2},\qd\r_Z,
     \qd\r_{Z_1}/\r_{Z_2},&\cr
     &\r_{X_1}/q_1,\qd\r_{X_1}\r_{Y_1}/q_1,\qd\r_{X_1}\r_{Y_1}\r_{Z_1}/q_1,\qd
     q_X,\qd q_Y,\qd q_Z,\qd q_1/q_2,\qd q_1,& \eea
we see that in this set $q_X$ has nontrivial DT only with $q_1$, so
we can eliminate these two generators. But then $q_Y$ has nontrivial DT only
with $\r_{X_1}/q_1$, so we also cancel these two generators.
In a third step we notice that now $q_Z$ has nontrivial DT only
with $\r_{X_1}\r_{Y_1}/q_1$, so after eliminating these two generators
we have reduced our set to
\beq \r_X,\qd \r_{X_1}/\r_{X_2},\qd\r_Y,\qd \r_{Y_1}/\r_{Y_2},\qd\r_Z,\qd
   \r_{Z_1}/\r_{Z_2},\qd \r_{X_1}\r_{Y_1}\r_{Z_1}/q_1,\qd q_1/q_2, \eeq
which is equivalent to a new basis given by
\beq \r_X,\qd \r_{X_3}\r_{Y_1}\r_{Z_1}/q_1,\qd\r_Y, \qd
    \r_{Y_3}\r_{X_1}\r_{Z_1}/q_1,\qd \r_Z,\qd \r_{Z_3}\r_{X_1}\r_{Y_1}/q_1,\qd
     \r_{X_1}\r_{Y_1}\r_{Z_1}/q_1,\qd q_1/q_2, \eeq
where we can cancel the last two generators.
Now all torsions come from the $q_1$'s, so this is equivalent to an
orbifold with a twist group $\ch$ generated by
\beq \r_X,\qd\r_Y,\qd\r_Z,\qd\r_{X_3}\r_{Y_1}\r_{Z_1},\qd
   \r_{Y_3}\r_{X_1}\r_{Z_1},\qd\r_{Z_3}\r_{X_1}\r_{Y_1}, \eeq
with trivial torsions among the first and last three generators
and torsions of exp$(4\pi i/3)$ between any of the first and any of the
last three generators.
In a matrix notation we thus find
{\small
\beq
     \hbox{\large$\ch\sim$} \pmatrix{ 1 & 1 & 1 & 0 & 0 & 0 & 0 & 0 & 0 \cr
     	0 & 0 & 0 & 1 & 1 & 1 & 0 & 0 & 0 \cr
	0 & 0 & 0 & 0 & 0 & 0 & 1 & 1 & 1 \cr
     0 & 0 & 1 & 1 & 0 & 0 & 1 & 0 & 0 \cr
     1 & 0 & 0 & 0 & 0 & 1 & 1 & 0 & 0 \cr
     1 & 0 & 0 & 1 & 0 & 0 & 0 & 0 & 1 \cr}
	,~~~~~~
     \hbox{\large$\e\sim$} \pmatrix{ 0 & 0 & 0 & 2 & 2 &2 \cr
	0 & 0 & 0 & 2 & 2 & 2 \cr 0 & 0 & 0 & 2 & 2 & 2 \cr
	1 & 1 & 1 & 0 & 0 & 0 \cr 1 & 1 & 1 & 0 & 0 & 0 \cr
	1 & 1 & 1 & 0 & 0 & 0 \cr}
\eeq}%
for the mirror of the model of (\ref H).
The spectrum of this
model is indeed again described by % the mirror of
(\ref{hd}), which is its own mirror spectrum.

\bigskip
{\it Acknowledgements.}
We would like to thank Bert Schellekens for his help in calculating the
singlet spectrum, and Albrecht Klemm and Christoph Schweigert for helpful
discussions. This work is supported in part by the
{\it "Osterreichische Nationalbank} under grant No. 5026.
M.K. acknowledges the hospitality of the CERN theory division.
% while this work was finished.


\newpage
\begin{thebibliography}{11}\let\bib=\bibitem
\addtolength{\itemsep}{-3pt}
\bib{sc90r}A.N.Schellekens and S.Yankielowicz, \ijmp 5 (1990) 2903 % 2/90
\bib{sci}M.Kreuzer and A.N.Schellekens, \npb 411 (1994) 97
\bib{gp}B.R.Greene and M.R.Plesser, \npb 338 (1990) 15
\bib{co}P.Candelas, X.C. de la Ossa, P.S.Green and L.Parkes,
    \npb359 (1991) 21;\\ \plb 258 (1991) 118
\bib{lvw}W.Lerche, C.Vafa and N.P.Warner, \npb 324 (1989) 427
\bib{mvw}E.Martinec, \plb 217 (1989) 431;\\
    C.Vafa and N.Warner, \plb 218 (1989) 51
\bib{bh}P.Berglund and T.H"ubsch, \npb 393 (1993) 377
\bib{eg}P.Berglund, M.Henningson, \npb 433 (1995) 311
\bib{mmi}M.Kreuzer, \plb 328 (1994) 312
\bib{dt}C.Vafa, \npb 273 (1986) 592
\bib{nms}M.Kreuzer and H.Skarke, \npb 388 (1992) 113;\\
	A.Klemm and R.Schimmrigk, \npb 411 (1994) 559
\bib{aas}M.Kreuzer and H.Skarke, \npb 405 (1993) 305
\bib{lgt}M.Kreuzer and H.Skarke, {\it Landau--Ginzburg orbifolds with discrete
	torsion,} \\preprint hep-th/9412033, ~to appear in Mod. Phys. Lett. A
\bib{toric}V.V.Batyrev, J. Alg. Geom. {\bf 3} (1994) 493
\bib{fiq}A.Font, L.E.Ib\'a\~nez, F.Quevedo, \plb217 (1989) 272
\bib{privat}L.E.Ib\'a\~nez, private communication
\bib{vw}C.Vafa and E. Witten, J. Geom. and Phys. {\bf 15} (1995) 189
\bib{viv}C.Vafa, \mpl 4 (1989) 1169; Superstring Vacua, HUTP-89/A057 preprint\\
	K.Intriligator and C.Vafa, \npb 339 (1990) 95
\bib{q}C.Vafa, \mpl 4 (1989) 1615
\bib{sy}A.N.Schellekens and S.Yankielowicz, \npb 330 (1990) 103;\\
    Tables Supplements CERN-TH.5440S/89 and CERN-TH.5440T/89 (unpublished)
\bib{ade}M.Kreuzer and H.Skarke, \plb 318 (1993) 305
% \bib{z2z2}A.E.Faraggi, \plb326 (1994) 62
% \bib{orbi}L.E.Ib\'a\~nez, J.Mas, H.P.Nilles, F.Quevedo, \npb 301 (1988) 157;
% 	A.Font, L.E.Ib\'a\~nez, F.Quevedo, \plb217 (1989) 272;\\%Z_MxZ_N+d.t.
% 	T.Mohaupt, MS-TPI-93-09 preprint
% 	% G.Lopes Cardoso, D. L"ust, T. Mohaupt, HUB-IEP-94-6, hep-th/9405002
% \bib{fiqs}A.Font, L.E.Ib\'a\~nez, F.Quevedo and A.Sierra, \npb 337 (1990) 119
\end{thebibliography}
\end{document}